# Mid-infrared directional surface waves on a high aspect ratio nano-trench platform

Osamu Takayama,[1] Evgeniy Shkondin,[1,2] Andrey Bodganov,[3] Mohammad Esmail Aryaee Panah,[1] Kirill Golenitskii,[4] Pavel Dmitriev,[3] Taavi Repän,[1] Radu Malureanu,[1] Pavel Belov,[3] Flemming Jensen,[2] and Andrei V. Lavrinenko[1]

[1]DTU Fotonik—Department of Photonics Engineering, Technical University of Denmark, Ørsteds Plads 343, DK-2800 Kgs. Lyngby, Denmark

[2]DTU Danchip—National Center for Micro- and Nanofabrication, Technical University of Denmark, Ørsteds Plads 347, DK-2800 Kgs. Lyngby, Denmark

[3]International Research Centre for Nanophotonics and Metamaterials, ITMO University, St. Petersburg, Russia

[4]Ioffe Institute, Russian Academy of Sciences, 194021, St. Petersburg, Russia

**Abstract**

Optical surface waves, highly localized modes bound to the surface of media, enable manipulation of light at nanoscale, thus impacting a wide range of areas in nanoscience. By applying metamaterials, artificially designed optical materials, as contacting media at the interface, we can significantly ameliorate surface wave propagation and even generate new types of waves. Here, we demonstrate that high aspect ratio (1 to 20) grating structures with plasmonic lamellas in deep nanoscale trenches, whose pitch is $1/10 - 1/35$ of a wavelength, function as a versatile platform supporting both surface and volume infrared waves. The surface waves exhibit a unique combination of properties, such as directionality, broadband existence (from 4 μm to at least 14 μm and beyond) and high localization, making them an attractive tool for effective control of light in an extended range of infrared frequencies.

**Main text**

Optical surface waves (SWs) arise at the interface of two dissimilar media with different types of permittivity or permeability[1]. Research on SWs has intensified in the last decade due to their unique properties of surface sensitivity, field localization, unusual dispersion and polarization properties at the nanoscale, stimulating the development of *surface photonics*[2]. The most studied SWs are surface plasmon-polaritons supported at the interfaces between metals and dielectrics[3], which enable effective nanophotonic devices for sensing[4], nano-guiding[5], and imaging[6] based on near–field techniques. A newly emerging alternative is Dyakonov surface waves existing at the interfaces between anisotropic and isotropic dielectrics[7-10]. Up to present, various types of SWs have mostly been investigated individually. However, we can obtain new features by combining traits from various types of surface waves. This is where metamaterials, an artificially engineered materials and structures[11-13], can play an essential role because in order to combine different SWs unprecedented and extreme optical parameters are often required. One example of such combined SWs on metamaterial structures are Dyakonov plasmons (DPs)[14,15], a combination of surface plasmons and Dyakonov waves supported at the boundaries of hyperbolic metamaterials (HMMs)[16]. The diagonal components of the HMMs' permittivity tensors are of different signs, giving rise to hyperbolic iso-frequency contours in the *k* (wavevector) space accompanied by singularities in the density of optical states in an ideal lossless case. Natural material equivalents of HMMs are often referred to as indefinite media[17,18]. Characteristically, HMMs and their two-dimensional analogues of metasurfaces possess a unique combination of properties including unusually high wavevectors, optical density of states, and anisotropy. These feature lead to a wide variety of HMM potential applications such as broadband enhancement in the spontaneous emission for a single photon source[19,20], sub-wavelength imaging[21], sensing[22,23], thermal engineering[19,20,24], and steering of optical signals[25,26]. Up to present, various types of optical HMMs have been fabricated and characterized, including LC circuit implementation for microwave frequency[25], metal-dielectric multilayers[23,27-29], shallow metallic gratings or metasurfaces[26,30,31], metallic pillar structures[22,32,33] and fishnet structures[24]. Most likely HMMs will be employed as the robust and versatile multi-functional photonic platform in the broad range of operating wavelengths from visible[26] to THz regions[34] and even at microwave region[25].

Currently, mid-infrared (IR) light with wavelengths between 2.5–25 μm (4,000 cm$^{-1}$ to 400 cm$^{-1}$) is used for thermal imaging and molecular detection (vibrational modes spectroscopy) [35,36]. In this wavelength range, DPs can enhance the applicability of the sensing devices thanks to their ability to localize light at the nanoscale. Due to the fact that noble metals are not suitable for confining light at interfaces on mid-IR frequencies, the quest for alternative materials has been on-going and a large variety of materials have been proposed, such as transparent conductive oxides[37], doped III-V semiconductors[35] and 2D materials [graphene, hexagonal boron nitride (h-BN), etc.][38-40].

Here, we realize a hybrid bi-slab metamaterial platform for mid-IR surface photonics, providing flexible engineering and adiabatic tuning of directivity and dispersion of highly localized and directional SWs, in particular DPs. No any conventional approaches in fabrication of HMM – neither multilayer stacks[23,27-29], nor wire medium[22,32,33] are able to bring pronounced anisotropy on the working interface of the platform. To attain such anisotropy we base the platform on a high-aspect ratio (up to 1:20 and even more) trench structures (Fig. 1) fabricated using a combination of atomic layer deposition (ALD) and dry etch techniques (see Method, Supplementary Fig. 1 and Supplementary Note 1). As a plasmonic constituent material required for the manifestation of effective hyperbolic dispersion we used aluminium-doped ZnO (AZO) exhibiting plasmonic response in the near and mid-IR wavelength region[37]. AZO can be deposited by the ALD technique[41] allowing a conformal and uniform coating of deep trenches. It can also benefit from the tunability of plasmonic properties by means of adjusting the doping concentration during the deposition. The period of trenches ($\Lambda = 0.4$ μm) is deeply subwavelength for the mid-IR wavelength range ($\Lambda/\lambda = 1/10 \sim 1/35$ for $\lambda = 4 \sim 14$ μm), allowing introduction of effective parameters. Controlling the etching time and rate accurately, we can reach the desired Si etching depth. Thus, three principal designs of HMMs can be obtained (Fig. 1b-d): AZO trenches embedded in Si (AZO/Si) or air (AZO/air) and a hybrid bi-slab metamaterial of AZO/air on top of AZO/Si. Consequently, the effective properties of the hybrid composite structure can be accurately tuned to exhibit hyperbolic, elliptic or epsilon-near-zero (ENZ) dispersion regimes with a plethora of ground-breaking applications including nonlinear optics[42,43]. To the best of our knowledge, this is the first experimental demonstration of directional surface (Dyakonov plasmons) and bulk (bulk plasmons) waves in the mid-infrared range on metamaterial platform with a high degree of flexible design parameters.

**Results**

    **Surface and bulk waves on air-plasmonic nano-trench hyperbolic metametamaterials.** In order to find the wavelength range, where the existence condition of DPs is met, we retrieve the effective permittivities (ordinary $\varepsilon_o$ and extraordinary $\varepsilon_e$) of the AZO/air structure from the mid-IR reflection spectra (see Method, Supplementary Figs. 2 – 4 and Supplementary Note 2). As shown in Fig. 2a, the AZO/Air trench structure has the ENZ regime for $\varepsilon_o$ around $\lambda_{ENZ} = 2.5$ μm and becomes Type II HMMs ($\varepsilon_o < 0$ and $\varepsilon_e > 0$)[19] within an extremely wide band at longer wavelengths. The existence condition for the DPs is $0 < \varepsilon_c < |\varepsilon_o|$[15], where $\varepsilon_c$ is the permittivity of the isotropic media bordering with the HMM (in our case, $\varepsilon_c = 1$, air). As highlighted in the inset of Fig. 2a, the condition is satisfied for $\lambda > 4.0$ μm.

    DPs are hybrid polarization waves and, therefore, can be excited by incident light of any polarization. We characterize directional waves in the AZO/Air trench structures in the mid-IR wavelength interval $\lambda = 2.0$ - $14.0$ μm (Fig. 1a and Fig. 2). The incident light reaches the structure through a hemispherical ZnSe prism (Fig. 1a) arranged in the Otto configuration. Reflection spectra are acquired under different angles of incidence and structure orientation. It allows mapping the reflectance in the $k_x$–$k_y$ space, see such maps for $\lambda = 6.0$ μm in Fig.2b,c. Zones with low reflectance outside the air light cone $k_x^2 + k_y^2 = k_0^2$, where $k_0 = \omega/c$, are potential candidates for SWs bands. To elucidate the nature of the optical modes population, we simulate the reflectance spectra of the AZO/Air structure in the Otto configuration (Fig.2e,f) and map the field profiles as shown in Fig.2d (see Methods). The reflectance spectra maps (Fig. 2b,c,e,f) obtained in the hyperbolic regime at $\lambda = 6.0$ μm clearly expose two bands. The profiles of the electric field amplitudes plotted in three characteristic points (A, B, C, Fig. 2e) clearly distinguish

the surface and volume waves bands (Fig. 2d). At point A, the fields are strongly confined at the interface between air and the HMM featuring a SW (Fig. 2d, panel A). The confinement of the SW is changed at higher in-plane wavenumbers (Fig. 2d, panel B), suggesting the unique feature to control the localization level of fields via SWs without changing wavelength. Plotting the field map in point C (Fig. 2d, panel C) enables us to identify a volume mode with the energy density concentrated predominantly inside the structure. The surface waves are definitely classified as DPs, and volume waves correspond to bulk plasmon-polaritons (BPPs)[27], which together represent the integral feature of the HMMs[16].

The thickness of the air gap between the prism and trench structure is evaluated by fitting the simulated reflection dips with the experimental ones, giving the best matching for the airgap of 0.5 μm (see details in Supplementary Fig. 5 and Supplementary Note 3). Checking the wavelength dependency we observe that DPs emerge clearly after $\lambda = 4.0$ μm exactly as predicted by effective parameters analysis in Fig. 2a (the condition is satisfied at $\lambda > 4.0$ μm) and extend to the high k-region for longer wavelengths up to $\lambda = 14.0$ μm. Thus, their broadband existence within the Type II hyperbolic region (see Supplementary Figs. 8 and 9) is confirmed. Importantly, the very good agreement between simulated and experimental data suggests a remarkable robustness of DPs and BPPs against inevitable fabrication and characterization imperfections.

**Surface and bulk waves on hybrid bi-slab hyperbolic metametamaterials.** The AZO/Air trench structure (Fig.1b) characterized above is the final product of the complete etch of interstitial Si between the AZO layers. In principle, the depth of Si etching $L$ is a free parameter, which defines the flexibility of the initial AZO/Si template (Fig.1d) towards adjusting it to a particular wavelength range. In Fig. 1c we show one intermediate example of the controllable Si etch with $L = 1.65$ μm. Such a composite system can be interpreted as consisting of two anisotropic trench structures of thicknesses $L$ (AZO/air) and $H$-$L$ (AZO/Si) placed one above another. We will refer to this configuration as the bi-slab model. Two extreme cases of the composition are the pure AZO/Si slab ($L = 0$) and pure AZO/air slab ($L = H$). In general the bi-slab structure cannot be homogenized completely, but the homogenization procedure is justified in application to each of the slab individually. So it is modelled as a composition of two serial homogeneous anisotropic slabs of thicknesses $L$ and $H$-$L$. In such case simulations of the bi-slab model can be conducted with a conventional transfer matrix method (see Supplementary Figs. 6 and 7, and Supplementary Note 4 and 5 for accuracy analysis).

We analyse changes in the properties of the bi-slab structure with gradual virtual growth of interstitial Si between the AZO layers. The evolution of the iso-frequency contours at $\lambda = 3.0$ μm with the gradual filling of the trenches by silicon is shown in Fig. 3a-d. We mark also the light cone for ZnSe (green line), Si (blue) and Ge (red) prisms, designating the areas within which modes can be excited in our experimental setup. For the better visualization of bands in Fig.3 the losses were reduced in 100 times, and data are presented for the TM polarization. Accordingly to estimations the structure supports propagation of directional high-k DPs and BPPs at $\lambda > 4.0$ μm, so we do not observe the signature of surface waves for $\lambda = 3.0$ μm in the deep etch case with $L = H$ (Fig.3a) corresponding to the AZO/air structure (Fig. 1b). Partial filling of the interstitial voids with Si arranges better conditions for bulk waves, as shown in Fig.3b (compare with experimental result in Supplementary Fig.12b). Further filling with Si up to $L = 0.5$ μm, makes the iso-frequency contour remarkably flat (Fig. 3c), leading to divergence-free propagation[44]. It is a consequence of the transition of the AZO/Si structure at wavelengths $\lambda = 2.5 - 3.5$ μm from Type II to Type I HMM, accompanied by drastic changes in the whole mode structure (see similar changes with Si filling illustrated in Supplementary Fig. 7 for $\lambda = 6.0$ μm). With $L = 0$ μm (Fig.3d) the case of a pure AZO/Si slab (Fig.1d) is reached. The hyperbolic modes become lossy due to the leakage into the substrate and transit to higher wavevectors range. Instead, the band with elliptic iso-frequency contour manifests itself close to the light cone.

In the lower panels (Fig. 3e-h) we analyse the transformation of the iso-frequency contours for $\lambda = 5.0$ μm, on which conditions for the DPs are satisfied. DPs bands are clearly visible in Fig.3e,f as the most intensive hyperbola-like curves (see the corresponding experimental data in Fig.8d and Fig.12d of Supplementary). Due to the exponential decay of the fields inside the structure for DPs, waves on the upper slab do not feel the presence of the lower one for rather extended range of sizes $L$. It is remarkably exhibited in Fig.3f with $L = 1.65$ μm, (Fig.3f), where the influence of the AZO/Si slab on the behaviour of DPs is not significant, although we see the drastically changed BPPs dispersion (see also Fig.7 in Supplementary). When the exponential tail of the DPs fields reaches the lower high-index slab, it affects dispersion of the SWs, resulting in the hybridization of the DPs and BPPs clearly visible as the band visibly distorts in Fig. 3f for $L = 0.5$ μm. Further filling with Si completely ruins conditions for the DPs existence (Fig. 3g), but promotes the settling of an elliptic BPP (Fig. 3h) in the AZO/Si composite.

To comprehend the investigation of the bi-layer system as a flexible platform for directional high-k waves we characterized the hybrid bi-slab ($L = 1.65$ μm) and single AZO/Si ($L = 0$) samples (Fig. 1c,d). The hybrid system satisfies the existence condition for SWs in a reduced wavelength range. For $\lambda = 5.0$ μm (see Fig. 4a, left row), the fields of SWs are located mainly in the AZO/air trench structure (Fig. 4b, panels A, B), thus exhibiting a hyperbolic dispersion. However, at $\lambda = 14.0$ μm (Fig. 4a, right row), confinement of the fields worsen (Fig. 4b, panel C), resulting in modification of the dispersion and a more complex field profile. This suggests that the DPs' propagation direction can be controlled efficiently over a large angular in-plane domain by varying the wavelength. The mechanism of the surface wave's dispersion modification is that for longer wavelengths the tail of the evanescent field of the SW residing on the AZO/air slab protrudes through the slab and starts to interact with the AZO/Si slab thereby leading to elliptic dispersion. Hence, in order to support DPs, we need a sufficiently deep trench structure, *e.g.* $L = 1.65$ μm for $\lambda = 5.0$ μm. In this regard, the hybrid bi-slab model allows us to tailor the spectrally-dependent directionality more efficiently than single AZO/air or AZO/Si structures separately. Such a system can exhibit Type I or Type II hyperbolic dispersion, and transition points for the hyperbolic behaviour together with the relevant ENZ regime can be configured for wavelengths in the range from 2.5 (Fig. 2a) to 7.5 μm (Fig. 4d). The accuracy of the design-tuning properties is granted by the mature technology of Si etching and consequently precise control of the air-silicon filling fraction in the interstitial spaces between AZO lamellas. In other words the effective properties can be accurately tuned by the height of the AZO/air trench structure positioned directly on top of the AZO/Si multilayer.

In the case of the original AZO/Si multilayer grating not undergoing any Si etching ($L = 0$), the elliptic dispersion of modes (Fig. 4c) is completely different from what we observed in the case of the AZO/air multilayer (see Fig. 2b,e). SWs exist for $\lambda > 10.0$ μm, where the structure is Type II HMM (Fig. 4d). The normalized wavevector or effective mode index of the SWs on the AZO/Si structure must exceed $(Re(\varepsilon_e))^{1/2} \sim 5$, which is higher than what can be reached with assistance of a ZnSe ($n = 2.4$) or even a Ge ($n = 4.0$)[8] prism. Therefore, the bands visible in the reflection spectra mapping (Fig.4a,c) are leaky bulk modes existing in anisotropic dielectric ($\lambda = 5.0$ μm) or hyperbolic ($\lambda = 14.0$ μm) metamaterials. Again we would like to emphasize the excellent correspondence of the experimental results in Fig.4 with numerical pictures from Fig.3 e-h despite of the heavily reduced losses in modelling.

**Discussion**

For the mid-IR wavelengths, some of two-dimensional materials are known to support surface waves and bulk modes, such as plasmons on graphene, hyperbolic phonon-polaritons in hBN, chiral plasmons on $MoS_2$[39,40]. Recently, hyperbolic surface phonon-polaritons, which are the phonon-polariton equivalent of Dyakonov plasmons, have been observed at the wedge of an topological insulator $Bi_2Se_3$ proved by electron energy loss spectroscopy (EELS)[45] and hBN by scanning near-field optical microscope (s-SNOM)[46]. However, due to the resonance character of phonon-polaritons the existence of most of these modes is restricted to narrow bands attributed to specific materials. The DPs on the trench structures are shown to exhibit broadband existence (from 4 μm to at least 14 μm and beyond), as well as unique properties such as directionality and controllable localization. The tuning of surface and bulk modes in the hybrid bi-slab HMM platform can be conducted by 1) tuning the material properties by adjusting the amount of doping Al concentration in ZnO, and 2) tuning the effective (metamaterials) properties by controlling the thickness of Air/AZO slab on Si/AZO structures in the hybrid bi-slab configuration. Such flexibility and broad parametric space in the optimization regime enable 1) tuning of the operational wavelength from mid-IR to THz ranges, and 2) tuning the direction of Dyakonov plasmons propagation. In principle, tuning of the optical properties of the trench structure may also be activated by electric gating of the AZO layers[47]. The demonstrated HMMs can be harnessed with 2D materials to hybridize optical modes from both parts since the broadband existence of DPs extends over all operating wavelengths of 2D materials in the mid-infrared region[39,40].

In conclusion, we characterize directional surface waves supported by the deep trench structures in the mid-infrared wavelength range. The structures are based on multiple high-aspect ratio (1:20) sub-wavelength AZO trenches embedded into the supporting Si substrate. Well-established Si etching technology is used to fabricate either a single slab with AZO/air trenches or a bi-slab with AZO/air trenches on top of AZO/Si trenches. All fabrication steps are supported by the large-scale CMOS compatible highly reproducible technological approaches. An AZO/air trench structure performs after homogenization as a broadband HMM. We observed two bands of directional waves classified further as DPs and BPPs according to the field confinement mapping. The etching depth of interstitial Si is a free parameter for tuning the performance of the whole bi-slab system. Existence conditions for different regimes of such directional waves can undergo very fine adjustment by tailoring the relative thicknesses of the combined trench structures. The broadband hyperbolic behaviour of the trench structure allows for the wavelength sweeping regime as an additional mechanism of the performance tuning. Our demonstration of SW behaviour in the deep trench structured platform can be conceptually extended to other wavelength regimes such as the near-infrared, THz and even the visible one by the choice of the relevant material platform. This could enable, for instance, the use of DPs with controllable dispersion for photonics applications as routing and switching of optical signals. Confinement of light close to the interface can also be effectively used for mid-IR spectroscopy to detect traces of analyte molecules via surface waves-enhanced sensing[36,48].

**Methods**

**Sample fabrication.** Al-doped ZnO (AZO) high aspect ratio trench structures were prepared by combining atomic layer deposition (ALD) and dry etch techniques. The use of ALD in combination with a sacrificial Si template is a novel way to create high aspect ratio structures of metal oxides. The process starts with the fabrication of a silicon trench template using deep reactive ion etching (DRIE) and the template is conformal-coated with an AZO film using ALD. The thickness of the coating should be at least half of the maximum distance between the Si trenches to fill the spacing between them. The top part of the AZO coatings can be removed by dry etching using Ar+ ion sputtering. Then, the silicon layers can be etched away selectively by a conventional $SF_6$ based isotropic Si dry etch process. Supplementary Figure 1 shows a schematic of the described fabrication flow. This approach enables us to fabricate high-quality optical metamaterials with high aspect ratio (20:1) on $2\times2cm^2$ or even larger areas.

**FTIR spectrometry and fitting.** The reflectance spectrum of a 100 nm thick AZO film deposited on a double side polished (DSP) silicon wafer is measured at 12° incident angle using a VERTEX 70 FTIR spectrometer from Bruker (Supplementary Fig. 2). The measurements were performed at five different points on the sample in order to get averaged data. The reflectance from the sample is calculated using the intensity transfer matrix method[49] and the Drude model for the permittivity of AZO. The calculated reflectance spectrum is then fitted to the measured one, using an algorithm based on the Levenberg-Marquardt method[50] in order to find the parameters of the Drude model (high-frequency dielectric constant, plasma frequency and damping) for AZO (Supplementary Table 4).

The similar procedure is carried out for both AZO/Si and AZO/air trench structures where samples are oriented so that the electric field is parallel to the trenches when measuring the reflectance of ordinary waves and perpendicular to the trenches when measuring the reflectance of extraordinary waves (Supplementary Fig. 3). Considering the Drude-Lorentz model for the ordinary permittivity and the Lorentz model for the extraordinary permittivity, we obtained the effective permittivities (Supplementary Table 5 and Supplementary Fig. 4).

**Theoretical analysis.** The reflectance maps in Figs. 2 - 4, and Supplementary Figs. 5 - 7 are calculated using the transfer matrix method for anisotropic media[51]. The components of the effective permittivity tensor are restored from the experiments on measurement of the reflectance spectra for different polarization of the incident wave. The profiles of electric field intensity in Figs. 2 and 4 are calculated using full-wave numerical simulations (FEM) with Comsol Multiphysics[52]. Due to translation symmetry of the structures along the trenches we simulated the model in 2D geometry. The simulation domain consists of a single unit cell with the Floquet boundary conditions. The incident wave is set through the port boundary condition. The infinite thickness of the substrate is simulated by adding a perfectly matched layer.

**Prism coupling experiment.** Our experimental setup is based on the Otto configuration [see Fig. 1(a)] mounted on the FTIR spectrometer (VERTEX 70, Bruker). A hemispherical ZnSe prism is placed on the sample with an unavoidable air gap between the prism and the trench structures. The ZnSe prism is used due to its high refractive index and transparency in the mid-IR region (transparency window is between 0.6 and 17.0 μm). The measurements were conducted in the wavelength range of $\lambda = 2.0 – 16.6$ μm ($5000 – 602$ cm$^{-1}$) for both TE- and TM-polarized incident light for the three different structures as shown in Supplementary Figs. 8-13. A wire grid polariser is used for controlling the incident polarisation. The input light from the thermal light source of the FTIR spectrometer is linearly-polarized either in TM polarization, the magnetic field in the x-y plane or in TE polarization, the electric field in the x-y plane. The beam is focused on the structure through a parabolic mirror and the ZnSe prism.

**Acknowledgements**

Authors thank Vladimir M. Shalaev, Alexandra Boltasseva and Thomas Søndergaard for fruitful discussion and valuable comments. This work was supported by Villum Fonden "Blokstipendium," and "DarkSILD project No. 11116" and Direktør Ib Henriksens Fond, Denmark.

**Author contributions**

O. T. and A. V. L. conceived the problem. E. S. developed and conducted the fabrication procedure with the support of R. M. and F. J.  O. T. and E. S. performed characterization of waves supported by the structures. A. B., K. G., P. D., and T. R. performed computational analyses and simulations. M. E. A. P. performed the permittivity retrieval process. All authors participated in interpretation of results and writing the manuscript.

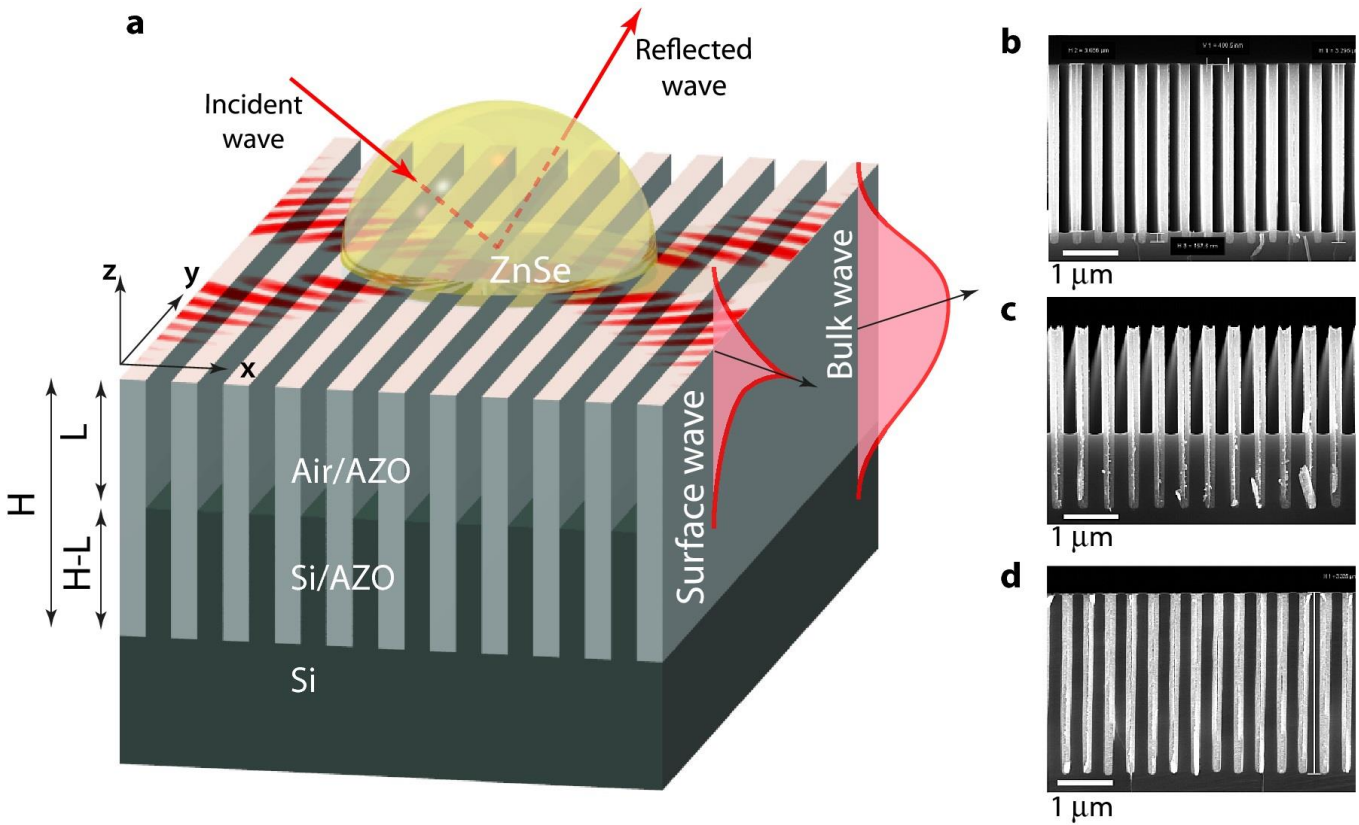

**Figure 1 | Hybrid bi-slab plasmonic trench structure.** (**a**) Illustration of an AZO trench structure in the Otto configuration based on a ZnSe prism for the FTIR spectrometer characterization. The structure drawn is a bi-layer of AZO/air trenches standing on AZO/Si ones on top of the Si substrate. Dyakonov plasmons and volume plasmon-polaritons can be supported by such vertical trench structures in the hyperbolic dispersion range. The optical axis of the homogenized metamaterial is perpendicular to the trenches. Scanning electron microscope (SEM) images of trench structures: (**b**) AZO/air, (**c**) AZO/air trench on top of AZO/Si trench, (**d**) AZO/Si. The structures have pitch $\Lambda$ = 400 nm, height of the entire trench structure $H$ = 2.8 – 3.3 μm, height of AZO/air trench $L$, filling ratio of AZO layer, $t_{AZO}/\Lambda$ = 0.5 where $t_{AZO}$ is the thickness of the AZO layer and are standing on the Si substrate. Note that the height of AZO/Si trench is $H - L$. The pitch of trenches is deeply subwavelength ($\lambda/10 - \lambda/35$) allowing introduction of effective parameters in the mid-IR range. The scale bars are 1μm for all figures. In this configuration, effective ordinary permittivity $\varepsilon_o$ and extraordinary permittivity $\varepsilon_e$ are oriented along, y- and z-axes, and x-axis, respectively.

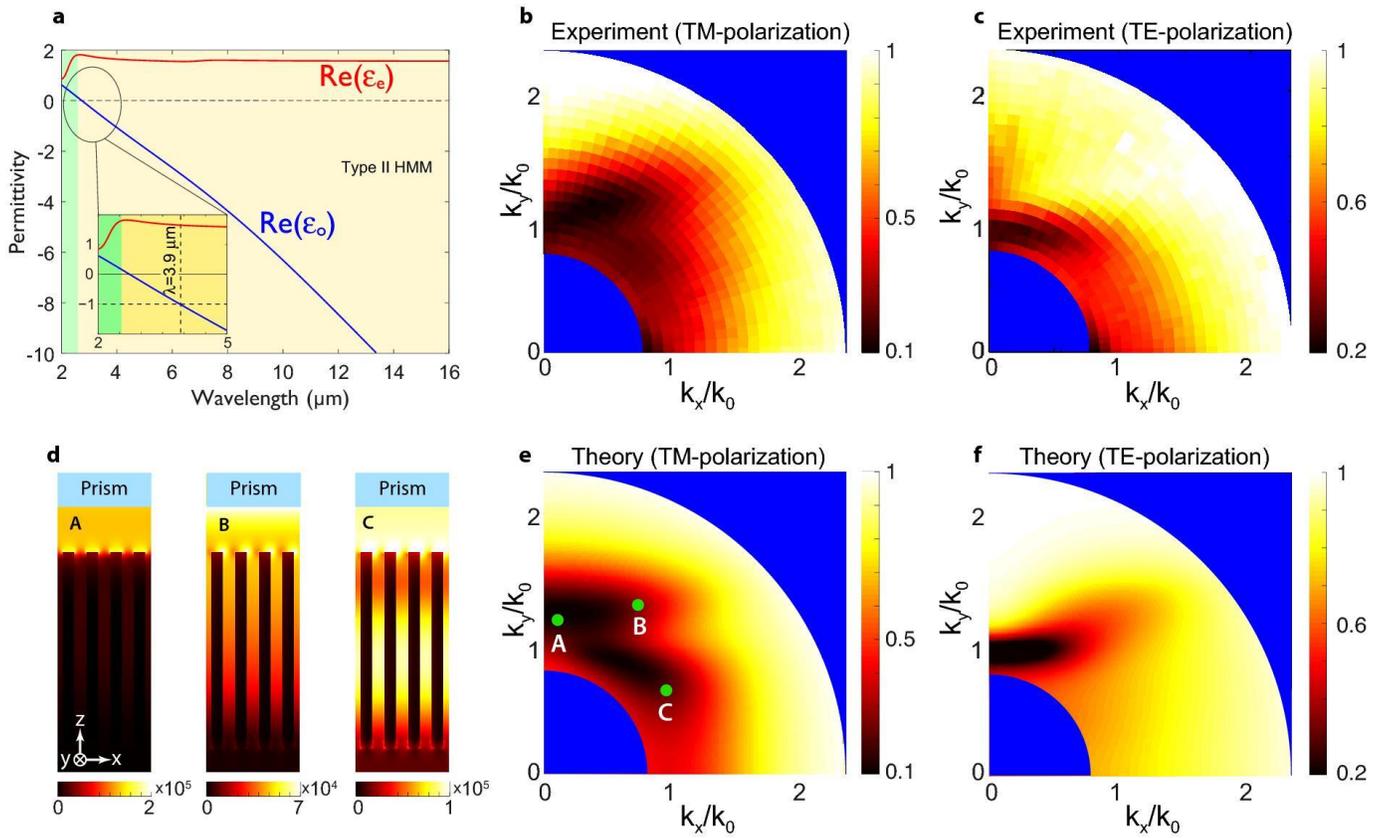

**Figure 2 | Observation of surface and bulk waves in AZO/air trench structure.** (**a**) Fitted real parts of effective ordinary $\varepsilon_o$ (blue) and extraordinary $\varepsilon_e$ permittivities (red) of an AZO/air trench metamaterial. The HMM range is marked in beige color. The inset shows in detail the restored real permittivities in the wavelength range 2 - 5 μm. The vertical dashed line in the inset designates the boundary wavelength for the DPs existence. (**b,c**) Experimental and (**e,f**) simulated reflectance in the wavevector space for $\lambda = 6.0$ μm with TM- and TE-polarized incident light. (**d**) Corresponding field profiles in points A, B and C from panel (**e**). The model includes the high index prism, air gap ($H_{air} = 0.5$ μm), trench structure ($L = 3.2$ μm AZO/air on top of $H - L = 0.1$ μm AZO/Si) and Si substrate.

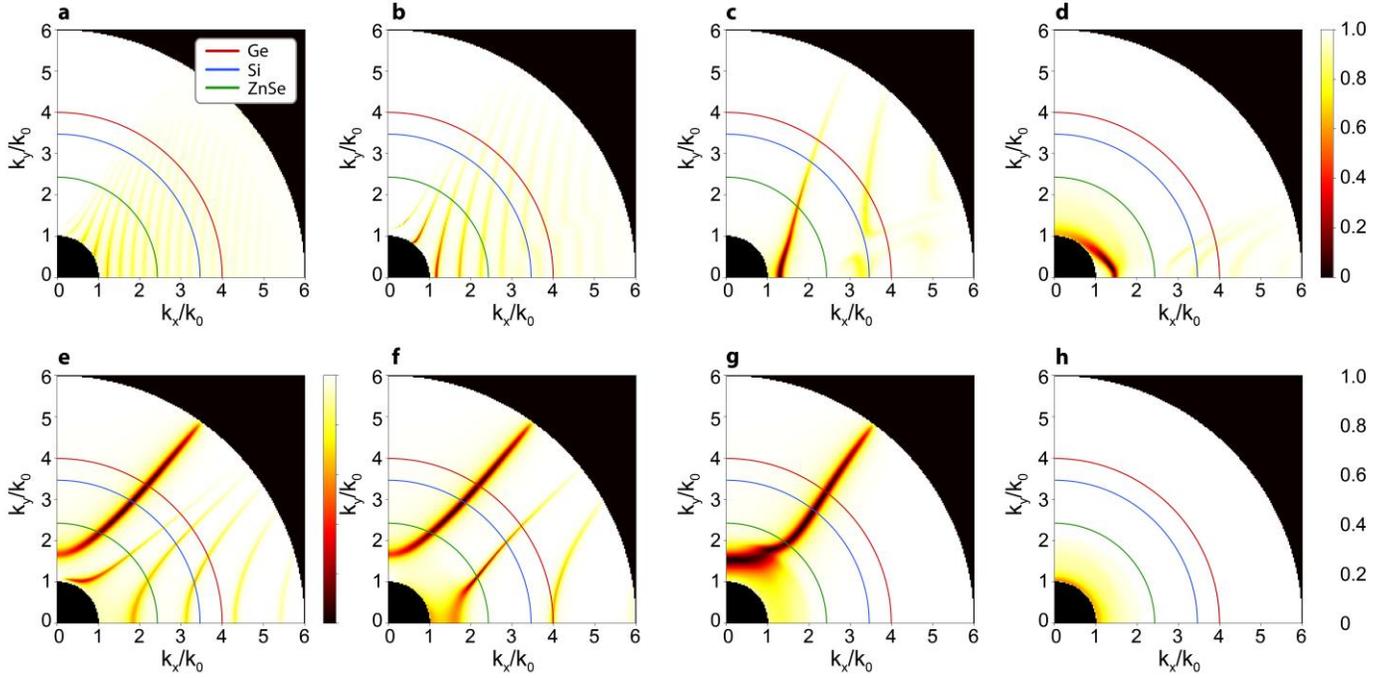

**Figure 3 | Transformation of band dispersion in the bi-slab structure for different etch depths of Si.** Band dispersion at $\lambda = 3.0$ μm, (**a**) $L = 2.8$ μm, (**b**) $L = 1.65$ μm, (**c**) $L = 0.5$ μm, (**d**) $L = 0$ μm; at $\lambda = 5.0$ μm, (**e**) $L = 2.8$ μm, (**f**) $L = 1.65$ μm, (**g**) $L = 0.5$ μm, (**h**) $L = 0$ μm, respectively. For the purpose of clarity in bands presentation, losses in the AZO layers are reduced by 100 times. Coloured quarter-circles designate the light cones for ZnSe (green), Si (blue) and Ge (red).

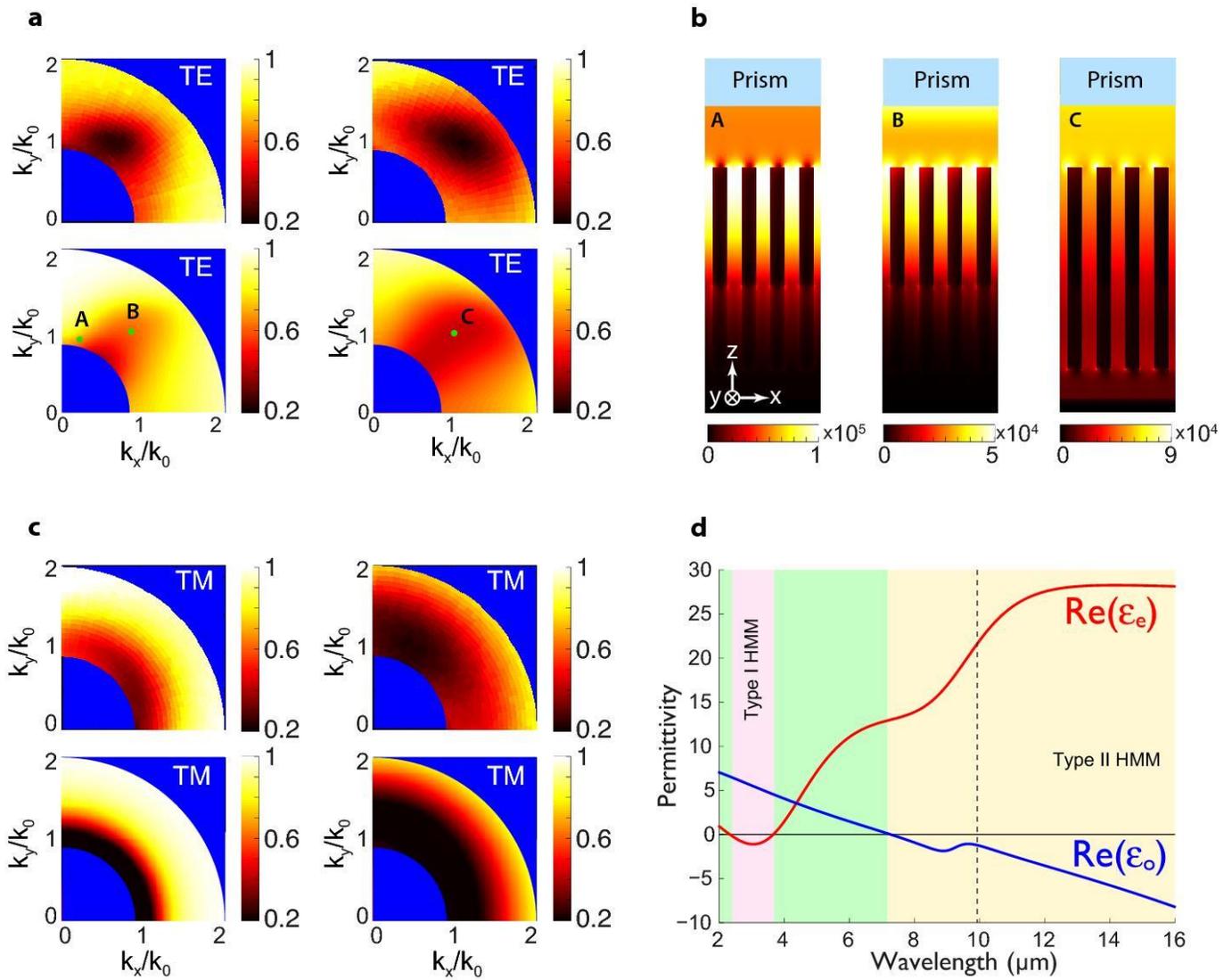

**Figure 4 | Surface waves on hybrid trench structures.** (**a**) Experimental and simulated reflectance in the wavevector space of the hybrid structure composed from the AZO/air ($L$ = 1.65 µm) layer on the AZO/Si layer ($H$-$L$ = 1.15 µm) for TE-polarized incidence light at $\lambda$ = 5.0 µm (left row) and, $\lambda$ = 14.0 µm (right row). (**b**) Corresponding electric field profiles at points A, B and C. (**c**) Experimental and simulated reflectance in the wavevector space of the AZO/Si multilayer ($L$ = 0 µm, $H$ = 3.2 µm) for TM-polarized light at $\lambda$ = 5.0 µm (left row), and $\lambda$ = 14.0 µm (right row). (**d**) Fitted real parts of the effective ordinary $\varepsilon_o$ (blue) and extraordinary $\varepsilon_e$ permittivities (red) for the AZO/Si ($L$ = 0 µm) trench structure. The vertical dashed line marks the boundary for the DPs range.